\documentclass[aps,a4paper,superscriptaddress,nofootinbib,twocolumn]{revtex4-1}
\usepackage{amsmath,amssymb,gensymb,graphicx,multirow,dcolumn,bm,latexsym,soul,nicefrac,acronym}
\usepackage[mathscr]{euscript}
\usepackage{appendix}
\usepackage[colorlinks,linkcolor=blue,citecolor=blue,urlcolor=blue]{hyperref}
\usepackage{float}
\usepackage{soul}
\usepackage{verbatim}
\usepackage{color}
\usepackage{acronym}
\usepackage{subfigure}
\usepackage{appendix}
\usepackage{longtable}
\usepackage{footnote}
\usepackage{cancel}
\usepackage{ulem} 

\newcommand{\mnras}{Monthly Notices of the Royal Astronomical Society}

\newcommand{\be}{\begin{equation}}
\newcommand{\ee}{\end{equation}}
\newcommand{\bea}{\begin{eqnarray}}
\newcommand{\eea}{\end{eqnarray}}
\newcommand{\bw}{\begin{widetext}}
\newcommand{\ew}{\end{widetext}}
\newcommand{\nn}{\nonumber}

\newcommand{\eq}[1]{Eq.~(\ref{#1})}
\newcommand{\fig}[1]{Fig.~\ref{#1}}
\newcommand{\tab}[1]{Table.~\ref{#1}}

\newcommand{\HNU}{School of Physics, Henan Normal University, Xinxiang 453007, China}

\newcommand{\TRC}{MOE Key Laboratory of TianQin Mission, TianQin Research Center for Gravitational Physics $\&$  School of Physics and Astronomy, Frontiers Science Center for TianQin, CNSA Research Center for Gravitational Waves, Sun Yat-sen University (Zhuhai Campus), Zhuhai 519082, China}

\begin{document}
\title{Inferring the stochastic gravitational-wave background from eccentric stellar-mass binary black holes with spaceborne detectors}
\author{Zheng-Cheng Liang}
\affiliation{\HNU}
\author{Zhi-Yuan Li}
\affiliation{\TRC}
\author{Yi-Ming Hu}
\email{huyiming@sysu.edu.cn}
\affiliation{\TRC}

\date{\today}

\begin{abstract}
The stochastic gravitational-wave background (SGWB) from eccentric stellar-mass binary black holes (SBBHs) holds crucial clues to their origins. 
For the first time, we employ a Bayesian framework to assess the detectability and distinguishing features of such an SGWB with spaceborne detectors, while accounting for contamination from the Galactic foreground. 
Our analysis covers eccentric SBBHs from three formation channels: isolated binary evolution, dynamical assembly in globular clusters (GCs), and in active galactic nuclei (AGNs). 
We find that TianQin, Laser Interferometry Space Antenna (LISA), and Taiji can detect the SGWBs from both isolated and GC-formed SBBHs after four years of operation, with the corresponding SNRs of around 10, 60, and 170. 
However, these backgrounds are spectrally degenerate with a strictly power-law SGWB. 
Furthermore, highly eccentric SBBHs formed in AGNs yield an SGWB marked by a spectral turnover and sharp decline. 
While this feature lowers the SNR by approximately an order of magnitude, it can enable a clear distinction from the strictly power-law background using LISA and Taiji. 
\end{abstract}



\maketitle
\acrodef{SGWB}{stochastic \ac{GW} background}
\acrodef{GW}{gravitational-wave}
\acrodef{CBC}{compact binary coalescence}
\acrodef{MBHB}{supermassive black hole binary}
\acrodef{BBH}{binary black hole}
\acrodef{SBBH}{stellar-mass binary black hole}
\acrodef{EMRI}{extreme-mass-ratio inspiral}
\acrodef{DWD}{double white dwarf}
\acrodef{GC}{globular cluster}
\acrodef{AGN}{active galactic nucleus}
\acrodef{LIGO}{Laser Interferometer Gravitational Wave Observatory}
\acrodef{TQ}{TianQin}
\acrodef{LISA}{Laser Interferometry Space Antenna}
\acrodef{TL}{TianQin + LISA}
\acrodef{KAGRA}{Kamioka Gravitational Wave Detector}
\acrodef{ET}{Einstein telescope}
\acrodef{DECIGO}{DECi-hertz interferometry GravitationalWave Observatory}
\acrodef{CE}{Cosmic Explorer}
\acrodef{NANOGrav}{The North American Nanohertz Observatory for Gravitational Waves}
\acrodef{LHS}{left-hand side}
\acrodef{RHS}{right-hand side}
\acrodef{ORF}{overlap reduction function}
\acrodef{ASD}{amplitude spectral density}
\acrodef{PSD}{power spectral density}
\acrodef{CSD}{cross-spectral density}
\acrodef{SNR}{signal-to-noise ratio}
\acrodef{TDI}{time delay interferometry}
\acrodef{PIS}{peak-integrated sensitivity}
\acrodef{PLIS}{power-law integrated sensitivity}
\acrodef{GR}{general relativity}
\acrodef{PBH}{primordial black hole}
\acrodef{SSB}{solar system baryo}
\acrodef{PT}{phase transition}
\acrodef{SM}{Standard Model}
\acrodef{EWPT}{electroweak phase transition}
\acrodef{CPTA}{Chinese Pulsar Timing Arrays}
\acrodef{PTA}{pulsar timing arrays}
\acrodef{RD}{radiation-dominated}
\acrodef{MD}{matter-dominated}
\acrodef{NG}{Nambu-Goto}

\section{Introduction}
The \ac{SGWB} originates from the superposition of numerous \acp{GW} that cannot be individually resolved~\cite{Christensen:2018iqi,Renzini:2022alw}. 
These persistent backgrounds can be broadly categorized as either astrophysical or cosmological in origin. 
The astrophysical SGWB is predominantly generated by populations of coalescing compact binaries, with its spectral characteristics encoding valuable information about their demographic properties and evolutionary pathways~\cite{Regimbau:2011rp,Bellomo:2021mer,Torres-Orjuela:2023hfd,Li:2024rnk}. 
On the other hand, the cosmological SGWB originates from fundamental physical processes in the early Universe, offering a precious observational window ~\cite{Nakayama:2008wy,Caldwell:2022qsj,Schulze:2023ich}.

Two methodologies have been proposed to extract the astrophysical information encoded in SGWBs: the cross-correlation method~\cite{Hellings:1983fr,1987MNRAS.227..933M,Christensen:1992wi,Flanagan:1993ix,Allen:1997ad} and the null-channel method~\cite{Tinto:2001ii,Hogan:2001jn,Smith:2019wny,Muratore:2021uqj}. 
The cross-correlation method, widely employed in current GW detectors, relies on the fundamental assumption that instrument noise between separated detectors is statistically independent. 
This enables the separation of a common SGWB signal from dominant noise contributions through cross-correlation of synchronized data streams. 
While ground-based detectors have yet to detect an SGWB in the frequency range of tens of hertz to kilohertz~\cite{LIGOScientific:2009qal,LIGOScientific:2011yag,LIGOScientific:2014gqo,LIGOScientific:2016jlg,LIGOScientific:2017zlf,LIGOScientific:2019vic,KAGRA:2021mth,KAGRA:2021kbb,LIGOScientific:2025bgj}, recent observations by \ac{PTA} have reported strong evidence for an SGWB in the nanohertz band~\cite{NANOGrav:2023gor,Xu:2023wog,EPTA:2023fyk,Reardon:2023gzh}. 
For future spaceborne detectors targeting the millihertz (mHz) band, including TianQin~\cite{TianQin:2015yph}, \ac{LISA}~\cite{LISA:2017pwj}, and Taiji~\cite{Ruan:2018tsw}, the null-channel method provides a promising detection alternative~\cite{Adams:2010vc,Adams:2013qma,Caprini:2019pxz,Boileau:2020rpg,Flauger:2020qyi,Banagiri:2021ovv,Cheng:2022vct,Boileau:2022ter,Gowling:2022pzb,Baghi:2023qnq,Wang:2022sti,Hartwig:2023pft,Muratore:2023gxh,Alvey:2023npw,Pozzoli:2023lgz,Li:2024lvt,Huang:2025uer}. 
This method utilizes a synthesized null channel that is intentionally insensitive to GWs while preserving the instrument noise characteristics. 
Through autocorrelation analysis of the detection channels combined with continuous noise monitoring via the null channel, the SGWB component can be effectively extracted without requiring a separate detector. 
Nevertheless, the performance of this method is inherently limited by uncertainties in detector noise characterization. 
Imperfect noise modeling can significantly bias the SGWB reconstruction, potentially degrading the accuracy of inferred background amplitudes by 1 to 2 orders of magnitude~\cite{Muratore:2023gxh}.

Among the dominant astrophysical sources, Galactic double white dwarfs can constitute a significant foreground~\cite{Bender:1997hs,Nelemans:2001hp,Barack:2004wc,Edlund:2005ye,Ruiter:2007xx,Nelemans:2009hy,Cornish:2017vip,Huang:2020rjf,Liang:2021bde,Liu:2023qap,Wang:2023jct,Staelens:2023xjn,Hofman:2024xar}, while \acp{SBBH} represent another major contributor to the SGWB. 
Conventional models of the SBBH-generated SGWB typically assume circular orbits when SBBHs enter the sensitive frequency bands of GW detectors~\cite{Zhu:2011bd,LIGOScientific:2016fpe,Inayoshi:2016hco,LIGOScientific:2017zlf,Yagi:2017zhb,Chen:2018rzo,Liang:2021bde}. 
This simplification can be well justified for the isolated evolution channel,  where binaries descend from massive stellar remnants~\cite{Belczynski:2001uc,Podsiadlowski:2002ww,Sadowski:2007dz,Kowalska:2010qg,Dominik:2012kk,Dominik:2013tma,Belczynski:2016obo,Breivik:2016ddj,Spera:2018wnw}. 
For such binaries, orbital eccentricities can be typically damped to values below $10^{-3}$ even at frequencies as low as $0.01\,\rm Hz$~\cite{Kremer:2018cir}. 
In contrast, this circular-orbit assumption becomes inadequate for SBBHs formed through dynamical processes in dense environments such as \acp{GC}~\cite{Banerjee:2009hs,Ziosi:2014sra,Rodriguez:2015oxa,Rodriguez:2016kxx,Banerjee:2016ths,Zevin:2018kzq,Fragione:2018vty,Rodriguez:2018pss,Kremer:2018cir,Martinez:2020lzt,DallAmico:2023neb,MarinPina:2025fnb,Dorozsmai:2025jlu} or \acp{AGN}~\cite{Antonini:2012ad,Hong:2015aba,Antonini:2016gqe,Hoang:2017fvh,Gondan:2017wzd,Leigh:2017wff,Takatsy:2018euo,Tagawa:2019osr,Samsing:2020tda,Tagawa:2020jnc,Rom:2023kqm}. 
In these environments, dynamical interactions can produce SBBHs with substantial eccentricities, with values exceeding $0.1$ persisting up to frequencies of $10\,\rm Hz$~\cite{Tagawa:2020jnc}.

The SGWB generated by eccentric SBBHs can exhibit a spectral shape that clearly differs from the strictly power-law spectrum generated by circular binaries~\cite{DOrazio:2018jnv,Zhao:2020iew,Xuan:2024dvx}. 
This distinction is particularly relevant in the mHz frequency band, where SBBHs remain in the early stages of inspiral.
Dynamically formed binaries in this regime are expected to retain significant orbital eccentricity due to their formation histories. 
Unlike circular binaries, which predominantly emit gravitational radiation at twice the orbital frequency, eccentric binaries can radiate across multiple harmonics of the orbital frequency. 
This broadband emission structure fundamentally alters the resulting energy spectral density $\Omega_{\rm gw}$, causing it to deviate from the characteristic $f^{2/3}$ power-law behavior.

In this paper, we investigate the SGWB produced by SBBHs from three distinct formation channels: isolated evolution, dynamical assembly in GCs, and in AGNs. 
To characterize spectral distortions induced by orbital eccentricity, we develop an analytical correction term that captures deviations from the strictly power-law spectrum. 
We begin with a preliminary detectability assessment for each SGWB with the TianQin, LISA, and Taiji detectors through \ac{SNR} calculation. 
We then employ a Bayesian framework that incorporates the Galactic foreground, which has been shown to significantly bias inferences of other SGWB components~\cite{Bender:1997hs,Nelemans:2001hp,Barack:2004wc,Edlund:2005ye,Ruiter:2007xx,Nelemans:2009hy,Cornish:2017vip,Huang:2020rjf,Liang:2021bde,Staelens:2023xjn,Hofman:2024xar,Liang:2024tgn,Pozzoli:2024wfe}. 
To improve computational efficiency, we derive a simplified likelihood function adapted to the null-channel method. 
Using this likelihood, we (i) perform Bayesian model selection to quantify evidence for an eccentric-SBBH-induced SGWB against contaminating Galactic foreground and instrument noise, and (ii) conduct parameter estimation to determine the precision with which LISA and Taiji can constrain the parameters characterizing the SGWB from eccentric SBBHs.

This paper is organized as follows: 
Sec.~\ref{sec:FD} presents the theoretical framework underpinning our analysis. 
Sec.~\ref{sec:method} details the statistical methodology for SGWB detection with spaceborne GW detectors. 
Sec.~\ref{sec:Results} presents and interprets our findings, including calculations of the eccentric-SBBH-induced SGWB, \ac{SNR} calculation, Bayesian model selection, and parameter estimation. 
Finally, we summarize our findings in Sec.~\ref{sec:Summary}, concluding with a discussion of the results and potential extensions.

\section{Theoretical foundations}\label{sec:FD}
\subsection{Statistical properties of stochastic background}
An \ac{SGWB} arises from the superposition of numerous unresolved GW sources. 
Since individual contributions cannot be distinguished, the characterization of such a background necessarily relies on statistical methods. 
In this work, we assume a Gaussian \ac{SGWB}, whose statistical properties are completely determined by its expectation value and variance.

The metric perturbations $h(t,\vec{x})$ related to an \ac{SGWB} can be represented as a superposition of plane waves propagating in all directions $\hat{k}$ and at different frequencies $f$~\cite{Misner:1973prb}, as follows:
\be
\label{eq:h_ab}
h(t,\vec{x})=\sum_{P}\int_{-\infty}^{\infty}{\rm d}f\int_{S^{2}}{\rm d}\hat{\Omega}_{\hat{k}}\,
\widetilde{h}_{P}(f,\hat{k})\textbf{e}^{P}(\hat{k}) e^{{\rm i}2\pi f[t-\hat{k}\cdot\vec{x}(t)/c]}.
\ee
Here, $\vec{x}$ denotes the detector position at time $t$, and $c$ is the speed of light. 
The polarization basis tensor $\textbf{e}^{P}$ spans the two transverse-traceless GW polarization states $P=+,\times$. 
The Fourier amplitude $\widetilde{h}_P$ encodes the frequency-domain properties. 
Under the additional assumptions that the SGWB is stationary, unpolarized, and isotropic, the statistical properties of the Fourier amplitudes satisfy the following relations:
\be
\langle\widetilde{h}_{P}(f,\hat{k})\rangle=0,
\ee
\be
\label{eq:Ph}
\langle\widetilde{h}_{P}(f,\hat{k})\widetilde{h}^{*}_{P'}(f',\hat{k}')\rangle
=\frac{1}{16\pi}\delta(f-f')\delta_{PP'}\delta^{2}(\hat{k},\hat{k}')S_{\rm h}(|f|),
\ee
where $\langle...\rangle$ denotes the ensemble average. 
The normalization factor $1/16\pi$ accounts for the two polarization states and integration over the celestial sphere, while $\delta$ and $\delta_{ij}$ represent the Dirac $\delta$ function and Kronecker $\delta$, respectively. 
The function $S_{\rm h}(f)$ corresponds to the one-sided \ac{PSD} of the background.

To quantify the GW energy density per logarithmic frequency interval normalized by the critical density of the Universe, 
a dimensionless energy spectral density $\Omega_{\rm gw}$ is commonly used. 
This parameter relates directly to the \ac{PSD} through~\cite{Thrane:2013oya}
\be
\label{eq:omega_gw}
\Omega_{\rm gw}(f)=\frac{1}{\rho_{\rm c}}\frac{{\rm d}\rho_{\rm gw}}{{\rm d}(\ln{f})}=\frac{2\pi^{2}}{3H_{0}^{2}}f^{3}S_{\rm h}(f),
\ee
where $\rho_{\rm gw}$ represents the \ac{GW} density, and $\rho_{\rm c}=3H_{0}^{2}c^{2}/(8\pi G)$ denotes the critical density, with $c$ being the speed of light, $G$ being the gravitational constant, and $H_{0}$ the Hubble constant.

\subsection{Detector channel and noise}\label{sec:s_n}
Spaceborne \ac{GW} detectors will employ a triangular constellation of three satellites optimized for mHz GW observations. 
TianQin will operate in a geocentric orbit with arm lengths of around $1.7\times10^{5}\,\rm{km}$ and a 3.64-day orbital period, implementing an intermittent observation schedule of ``three months on" alternating with ``three months off"~\cite{TianQin:2015yph}. 
Unlike TianQin, the LISA constellation will follow a heliocentric orbit, trailing Earth by approximately $20\degree$ while maintaining a $60\degree$ inclination to the ecliptic plane, with satellite separations of about $2.5\times10^{6}\,\,{\rm km}$~\cite{LISA:2017pwj}. 
In a geometrically complementary configuration, Taiji will be positioned symmetrically opposite to LISA relative to Earth, employing longer arm lengths of $3\times10^{6}~\mathrm{km}$~\cite{Ruan:2018tsw}.

The motion of spaceborne detectors introduces inherent phase noise that cannot be fully canceled. 
To suppress this noise below the level of target \ac{GW} signals, \ac{TDI} techniques have been proposed~\cite{Tinto:1999yr,Tinto:2020fcc} to produce effectively equal arm length interferometers that cancel the laser phase noise. 
The triangular configuration can provide six laser links that enable the construction of primary TDI channels, conventionally designated as $X$, $Y$, and $Z$. 
Each channel is centered on a specific satellite and synthesized from two laser beams traversing four distinct links. 
For example, denoting the three satellites as 1, 2, and 3, and representing the laser link from satellite 1 to 2 as $1\rightarrow2$, the $X$ channel combines two optical paths as follows:
\begin{equation}
	[1\rightarrow 2 \rightarrow1\rightarrow3\rightarrow1]-[1\rightarrow 3 \rightarrow1\rightarrow2\rightarrow1].
\end{equation}
The $Y$ and $Z$ channels are constructed through cyclic permutation of the satellite indices, maintaining identical topological structures but centered on different satellites.

Building upon the ($X$, $Y$, $Z$) basis, an advanced set of noise-orthogonal TDI channels ($A$, $E$, $T$) can be constructed through the linear combinations~\cite{Vallisneri:2007xa}, as follows:
\bea
\label{eq:channel_AET}
\nn
A&=&\frac{1}{\sqrt{2}}(Z-X),\\
\nn
E&=&\frac{1}{\sqrt{6}}(X+Y+Z),\\
T&=&\frac{1}{\sqrt{3}}(X+Y+Z).
\eea
Here, the $A$ and $E$ channels are primarily used for GW detection, while the $T$ channel serves as a null channel mainly employed to monitor instrument noise.

Assuming stationary noise, the noise \acp{PSD} of these channels are given by
\bea
\label{eq:Pn_AET}
\nn
P_{{\rm n}_{A/E}}(f)
&=&
\frac{2\sin^{2}u}{L^{2}}
\bigg[\big(\cos u+2\big)S_{\rm p}(f)\\
\nn
&&+2\big(\cos(2 u)+2\cos u+3\big)
\frac{S_{\rm a}(f)}{(2\pi f)^{4}}\bigg],\\
P_{{\rm n}_{T}}(f)
&=&
\nn
\frac{8\sin^{2}u
	\sin^{2}\frac{u}{2}}{L^{2}}
\bigg[S_{\rm p}(f)+4\sin^{2}\frac{u}{2}
\frac{S_{\rm a}(f)}{(2\pi f)^{4}}\bigg],\\
\eea
where $u=(2\pi fL)/c$ is a dimensionless parameter dependent on the arm length $L$. 
$S_{\rm p}$ and $S_{\rm a}$ denote the \acp{PSD} of optical-metrology system noise and acceleration noise, respectively. 
For comprehensive parameter specifications of LISA, Taiji, and TianQin, we refer readers to Refs.~\cite{Robson:2018ifk,Ren:2023yec,Li:2023szq}. 
Unless otherwise stated, this work primarily utilizes the $A/E/T$ channels for all subsequent analyses.

\subsection{Response to gravitational waves}\label{sec:response}
In the SGWB detection, the observational data comprise not only instrument noise but also the SGWB signal, which incorporates the detector response to GWs. 
The motion of the detector can introduce time-dependent variations in this response, while the direct processing of data collected over multiple years from the spaceborne detector demands prohibitively high computational resources. 
To improve computational efficiency, the data can be divided into shorter segments, and data folding can be applied~\cite{Ain:2015lea,Ain:2018zvo,Liang:2024tgn}. Each segment is centered at a time $t$ and spans the interval $[t-\tau/2, t+\tau/2]$, and is then analyzed using the short-time Fourier transform. 
The segment duration must be carefully chosen: it should be sufficiently long to resolve the relevant frequency bands after time-frequency conversion, yet short enough that the detector response remains approximately stationary over the interval. 
Within such a segment of duration $\tau$, the frequency-domain SGWB signal is given by the convolution of the impulse response $\mathbb{F}$ and the metric perturbation $h$~\cite{Romano:2016dpx}, as follows:
\bea
\label{eq:ht_sgwb}
\nn
h_{I}(t,\tau)&=&\mathbb{F}_{I}[t,\vec{x}_{I}(\tau)]*h[t,\vec{x}_{I}(\tau)]\\
\nn
&=&
\int_{-\infty}^{\infty}{\rm d}f \, \sum_{P}
\int_{S^{2}}{\rm d}\hat{\Omega}_{\hat{k}}
F_{I}^{P}(f,\hat{k},\tau)\widetilde{h}_{P}(f,\hat{k})\\
&&\times 
e^{{\rm i}2\pi f[t-\hat{k}\cdot\vec{x}_{I}(\tau)/c]},
\eea
where the frequency-domain response $F_{I}^{P}$ depends on the polarization tensor and the specific detector channel $I$~\cite{Cornish:2001qi}. 
In this framework, the \ac{PSD} and cross-spectral density of the frequency-domain signal $\widetilde{h}_{I,J}$ satisfy
\be
\label{eq:hIhJ}
\langle\widetilde{h}_{I}(f,\tau)\widetilde{h}_{J}^{*}(f',\tau)\rangle
=\frac{1}{2}\delta(f-f')\Gamma_{IJ}(f,\tau)S_{\rm h}(|f|),\\
\ee
where the \ac{ORF} incorporates both the channel response $F_{I}^{P}$ and the separation vector $\Delta \vec{x}=\vec{x}_{I}-\vec{x}_{J}$ between the two channels\footnote{In spaceborne detectors, a satellite may serve as the reference location for a channel.}~\cite{Liang:2022ufy}, as follows:
\bea
\label{eq:Gamma_IJ}
\nn
\Gamma_{IJ}(f,\tau)=&&
\frac{1}{8\pi}\sum_{P}
\int_{S^{2}}{\rm d}\hat{\Omega}_{\hat{k}}\,
F^{P}_{I}(f,\hat{k},\tau)F^{P*}_{J}(f,\hat{k},\tau)\\
&&\times
e^{-{\rm i}2\pi f\hat{k}\cdot \Delta \vec{x}(\tau)/c}.
\eea

\begin{figure}[h]
	\centering
	\includegraphics[height=6.6cm]{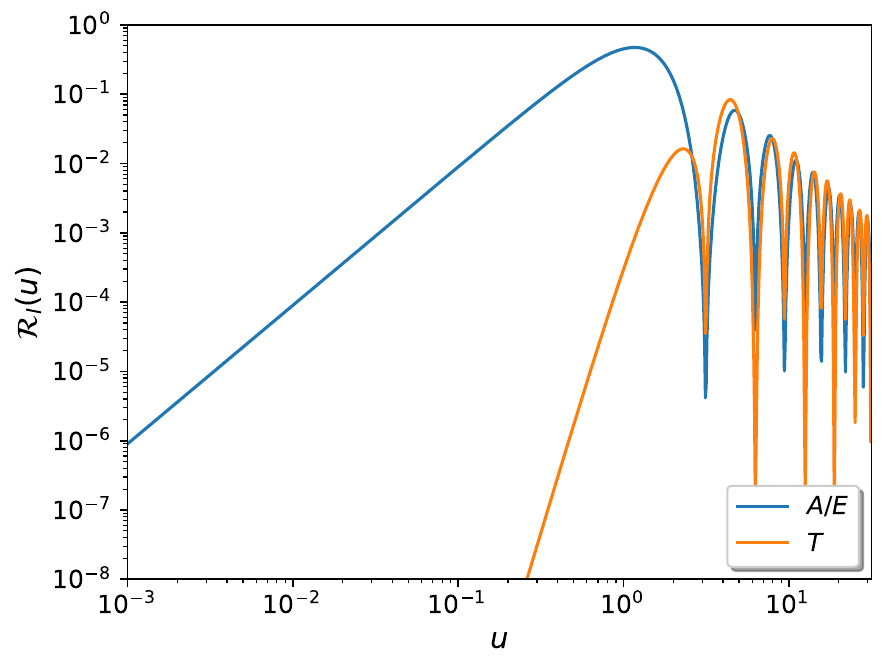}
	\caption{Transfer functions for the TDI channels as a function of the dimensionless frequency $u = (2\pi fL)/c$. The blue and orange lines correspond to the $A/E$ and $T$ channels, respectively.}
	\label{fig:ORF_AET}
\end{figure}

For a single detector channel, the \ac{ORF} simplifies to a transfer function $\mathcal{R}(f)$, which is fundamentally determined by the configuration and arm length of the detector~\cite{Cornish:2001qi}. 
For spaceborne detectors, which are typically arranged in an equilateral triangular configuration, the arm length serves as the key distinguishing parameter. 
To facilitate consistent comparisons across different detectors, we adopt the dimensionless frequency parameter $u$, which effectively normalizes the frequency dependence with respect to arm length. 
Fig.~\ref{fig:ORF_AET} illustrates the resulting transfer functions for different channels. 
Owing to the intrinsic symmetry of the detector configuration, the transfer functions for the $A/E$ channels are identical. 
In terms of spectral behavior, the $A/E$ channels scale as $f^2$, whereas the $T$ channel scales as $f^8$. 
This behavior leads to a significantly lower response in the $T$ channel compared to the $A/E$ channels at low frequencies. 
Nevertheless, as the frequency increases, the response of the $T$ channel rises sharply, eventually allowing it to contribute to GW detection.

\subsection{Sensitivity}\label{sec:sensitivity}
The sensitivity of a detector channel to an SGWB can be characterized by a sky-averaged sensitivity curve, defined using the noise PSD and the ORF as~\cite{Cornish:2001bb}
\be
\label{eq:S_nI}
S_{{\rm n}_{I}}(f)=\frac{P_{{\rm n}_{I}}(f)}{\mathcal{R}_{I}(f)}.
\ee
The associated noise energy spectral density is given by
\be
\label{eq:Omega_n}
\Omega_{{\rm n}_{I}}(f)=
\frac{2\pi^{2}}{3H_{0}^{2}}f^{3}S_{{\rm n}_{I}}(f).
\ee

In the weak-signal limit, the total \ac{SNR} from the $A$, $E$, and $T$ channels of a spaceborne detector can be written as~\cite{Caprini:2015zlo,Liang:2024ulf}
\be
\label{eq:snr}
\rho
=\sqrt{\sum_{I=A,E,T} T_{\rm ob}\int_{f_{\rm min}}^{f_{\rm max}}{\rm d}f\,
\bigg[\frac{\Omega_{{\rm gw}}(f)}{\Omega_{{\rm n}_{I}}(f)}\bigg]^{2}},
\ee
where $T_{\rm ob}$ denotes the total observation time. 
\eq{eq:snr} emphasizes the significance of both the observational duration and the accessible frequency band of the detector in SGWB detection.

The energy spectral density of an SGWB often follows a strictly power-law spectral form:
\be
\label{eq:Omegaform}
\Omega_{\rm gw}(f)=\Omega_{0}(\epsilon)(f/f_{\rm ref})^{\epsilon}|_{\epsilon=\epsilon_{0}},
\ee
where $\Omega_0$ depends on the spectral index $\epsilon$, and $f_{\rm ref}$ is a reference frequency. 
The amplitude $\Omega_0(\epsilon)$ required to achieve a certain SNR threshold $\rho_{\rm th}$ is given by
\be
\Omega_{0}(\epsilon)|_{\rho_{\rm th}}=\rho_{\rm th}\left[\sum_{I=A,E,T}T_{\rm tot}
\int_{f_{\rm min}}^{f_{\rm max}}{\rm d}f
\frac{(f/f_{\rm ref})^{2\epsilon}}{\Omega_{{\rm n}_{I}}^{2}(f)}\right]^{-1/2}.
\ee
The \ac{PLIS} curve is then constructed by taking the envelope of all such power-law spectra that yield an SNR equal to $\rho_{\rm th}$~\cite{Thrane:2013oya}, as follows:
\be
\label{eq:Omega_PLI}
\Omega_{\rm PLIS}(f)|_{\rho_{\rm th}}={\rm max}_{\epsilon}[\Omega_{0}(\epsilon)|_{\rho_{\rm th}}(f/f_{\rm ref})^{\epsilon}].
\ee
When visualized on a log-log scale, the PLIS curve delineates the detection threshold: spectra lying above the curve are detectable at the target SNR, while those falling below remain undetectable. 
This representation offers an intuitive and visually straightforward method for assessing the detectability of candidate SGWB signals.

\section{Statistical methodology}\label{sec:method}
\subsection{Bayesian inference}
Statistical inference offers a principled framework for identifying the presence of an SGWB in observational data. 
Both frequentist and Bayesian methods are widely employed in such analyses. 
In this work, we concentrate on Bayesian inference, which is particularly well-suited for model selection and parameter estimation.

Bayesian inference provides a rigorous probabilistic approach for quantifying uncertainties in unknown parameters. 
The process centers on constructing the posterior probability distribution via Bayesian analysis, which updates prior knowledge of the model parameters $\boldsymbol{\theta}$, encoded in the prior distribution $p(\boldsymbol{\theta})$, using information from the data through the likelihood function $p(d|\boldsymbol{\theta})$. 
The posterior distribution can be expressed as:
\be
\label{eq:pot_dib}
p(\boldsymbol{\theta}|d)=
\frac{p(d|\boldsymbol{\theta})p(\boldsymbol{\theta})}{p(d)},
\ee
where the normalization factor, known as the Bayesian evidence, is given by:
\be
p(d)=\int {\rm d \boldsymbol{\theta}}\,
p(d|\boldsymbol{\theta})p(\boldsymbol{\theta}).
\ee

Beyond parameter estimation, Bayesian inference provides a formal framework for model comparison. Suppose we have a set of candidate hypotheses $\mathcal{H}_{a}$, each characterized by a set of model parameters $\boldsymbol{\theta}_{a}$. 
The posterior probability of the model parameters under a given hypothesis is expressed as:
\be
p(\boldsymbol{\theta}_{a}|d,\mathcal{H}_{a})=
\frac{p(d|\boldsymbol{\theta}_{a},\mathcal{H}_{a})p(\boldsymbol{\theta}_{a}|\mathcal{H}_{a})}{p(d|\mathcal{H}_{a})},
\ee
where $p(d|\mathcal{H}_{a})$ denotes the evidence for hypothesis $\mathcal{H}_{a}$. 
The posterior probability of the hypothesis itself is given by
\be
p(\mathcal{H}_{a}|d)=\frac{p(d|\mathcal{H}_{a})p(\mathcal{H}_{a})}{P(d)},
\ee
with $P(d)$ representing the total evidence over all candidate hypotheses.
To compare two competing hypotheses $\mathcal{H}_{a}$ and $\mathcal{H}_{b}$, we can compute the posterior odds ratio:
\be
\mathcal{O}_{ab}(d)=\frac{p(\mathcal{H}_{a}|d)}{p(\mathcal{H}_{b}|d)}=\frac{p(\mathcal{H}_{a})}{p(\mathcal{H}_{b})}\frac{p(d|\mathcal{H}_{a})}{p(d|\mathcal{H}_{b})}.
\ee
In the absence of prior preference for either model, this ratio reduces to the Bayes factor,
\be
\mathcal{B}_{ab}=\frac{p(d|\mathcal{H}_{a})}{p(d|\mathcal{H}_{b})}.
\ee
The Bayes factor serves as a key metric for model selection. 
A value of $\ln (\mathcal{B}_{ab})>1$ indicates positive evidence in favor of $\mathcal{H}_{a}$, while $\ln (\mathcal{B}_{ab})>3$ and $\ln (\mathcal{B}_{ab})>5$ signify strong and very strong support, respectively~\cite{Robert:BF}.

To compute the model evidence and sample from the posterior distribution, we employ nested sampling~\cite{dynested}, a Monte Carlo method particularly suited for efficient evidence estimation and parameter inference. 
The algorithm begins by sampling a set of live points from the prior distribution~\cite{Skilling:2006gxv, Higson:2018cqj, Speagle:2019ivv, Ashton:2022grj}. 
At each iteration, the algorithm removes the point with the lowest likelihood and replaces it with a new point drawn from the prior, under the constraint that the new point has a higher likelihood. 
The iterative process continues until a predefined convergence criterion is met, with the dynamic allocation of samples ensuring thorough exploration even in challenging parameter spaces. 
This approach ensures progressive sampling toward higher-likelihood regions while maintaining detailed balance. 
Furthermore, the process naturally produces a set of weighted samples that can be used for both accurate evidence computation and parameter estimation, enabling efficient Bayesian model comparison and posterior inference within a single sampling run.

\subsection{Simplified likelihood}\label{subsec:llh}
In this section, we derive the specific form of the likelihood used in our analysis. 
Instead of processing the full multiyear data stream directly, we begin by dividing it into $N$ segments with equal duration $T$. 
For the $n$th segment centered at time $t_{n}$, the Fourier transform of the time-domain signal in channel $I\in \{A,E,T\}$ is given by
\be
\widetilde{s}^{n}_{I}(f)=\int_{t_{n}-T/2}^{t_{n}+T/2}{\rm d}t \,
s_{I}(t)e^{-{\rm i}2\pi ft}. 
\ee
Assuming the noise is Gaussian and uncorrelated across frequencies, the covariance matrix of the Fourier-domain data vector $\widetilde{\bm{s}}^{n}=(\widetilde{s}^{n}_{A}, \widetilde{s}^{n}_{E}, \widetilde{s}^{n}_{T})$ is diagonal,
\be
\mathcal{C}(\boldsymbol{\theta}, f) = 
\begin{pmatrix}
	P_{{\rm s}_{A}}(f) & 0 & 0 \\
	0 & P_{{\rm s}_{E}}(f) & 0 \\
	0 & 0 & P_{{\rm s}_{T}}(f)
\end{pmatrix},
\ee
where the total \ac{PSD} for each channel is defined as
\be
P_{{\rm s}_{I}}(f)=\mathcal{R}_{I}(f)S_{\rm h}(f) + P_{{\rm n}_{I}}(f).
\ee
The log-likelihood for the $n$th segment is then given by~\cite{Boileau:2020rpg}
\bea
\nn
\ln\left[p(\widetilde{\bm{s}}^{n}(f)|\boldsymbol{\theta})\right]
&=&-\frac{1}{2}\sum_{f}\sum_{I,J=A,E,T}
\big[\widetilde{s}^{n}_{I}(f)\mathcal{C}^{-1}(\boldsymbol{\theta}, f)\widetilde{s}^{n*}_{J}(f)\\
\nn
&&+\ln \left(2 \pi \det(\mathcal{C})\right)
\big]
\\
&=&-\frac{1}{2}\sum_{f,I}
\left[\frac{|\widetilde{s}^{n}_{I}(f)|^{2}}{P_{{\rm s}_{I}}(f)}
+\ln \left(2 \pi P_{{\rm s}_{I}}(f)\right)\right].
\eea
For the full dataset, the joint log-likelihood is obtained by summing over all segments,
\bea
\label{eq:llk}
\nn
\ln\left[\mathcal{L}(\boldsymbol{\theta})\right]
&=&\sum_{i}\ln\left[p(\widetilde{\bm{s}}^{n}(f)|\boldsymbol{\theta})\right]\\
&=&
-\frac{1}{2}\sum_{f,I}
\left[\frac{\mathcal{S}_{I}(f)}{P_{{\rm s}_{I}}(f)}
+\ln \left(2n \pi  P_{{\rm s}_{I}}(f)\right)\right],
\eea
where
\be
\mathcal{S}_{I}(f)=\sum_{n=1}^{N}|\widetilde{s}^{n}_{I}(f)|^{2}.
\ee
By dividing the full dataset into $N$ segments, the number of frequency points used in the likelihood evaluation is effectively reduced by a factor of $N$, significantly enhancing computational efficiency in subsequent Bayesian inference, without compromising statistical information. 
Note that the above derivation relies on the assumptions that both the noise and signal are stationary and that all data segments share the same duration.

\section{Results}\label{sec:Results}
\subsection{Stochastic background from eccentric stellar-mass binary black holes}\label{sec:BG}
SBBHs have been firmly observed by ground-based \ac{GW} detectors~\cite{LIGOScientific:2018mvr,LIGOScientific:2021usb,LIGOScientific:2021djp,LIGOScientific:2025slb}. 
During their early inspiral phase, such binaries emit GWs in the mHz frequency band, which lies within the observational range of spaceborne detectors~\cite{Sesana:2016ljz,Hu:2017yoc,Liu:2020eko,Zhang:2024fka}. 
The collective emission from a population of these binaries is expected to constitute an SGWB. 

For an ensemble of SBBHs described by population parameters $\boldsymbol{\lambda}$, such as component masses, spins, and orbital eccentricity, the energy spectral density of the resulting SGWB can be expressed as the integrated contribution of individual sources~\cite{LIGOScientific:2016fpe,LIGOScientific:2017zlf}, as follows:
\be
\label{eq:omegastro}
\Omega_{\rm gw}(f)=\frac{f}{\rho_{\rm c}}
\int\,{\rm d}\boldsymbol{\lambda}\int_{0}^{z_{\rm max}}\,{\rm d}z
\frac{R_{\rm m}(z,\boldsymbol{\lambda})\frac{{\rm d}E_{\rm gw}}{{\rm d}f_{\rm s}}(f_{\rm s},\boldsymbol{\lambda})}{(1+z)H(z)},
\ee
where $R_{\mathrm{m}}(z,\boldsymbol{\lambda})$ is the merger rate, and $\frac{\mathrm{d}E_{\mathrm{gw}}}{\mathrm{d}f_{\mathrm{s}}}(f_{\mathrm{s}}, \boldsymbol{\lambda})$ denotes the energy spectrum emitted in the source frame at frequency $f_{\mathrm{s}} = (1+z)f$. 
The Hubble parameter $H(z)$ can be approximated as $H_0 \sqrt{\Omega_{\mathrm{m}}(1+z)^3 + \Omega_{\Lambda}}$ up to $z_{\mathrm{max}} = 10$~\cite{Planck:2018vyg}, beyond which star formation, and consequently the formation of astrophysical black holes, becomes negligible.

In this work, we model the merger rate as redshift dependent, adopting a scaled cosmic star formation rate density~\cite{Madau:2014bja}, normalized to a local merger rate of $R_{\rm m}(0) = 19^{+7}_{-5}\,{\rm{Gpc^{-3}\,\,yr^{-1}}}$~\cite{LIGOScientific:2025pvj}. 
For the mHz frequency band, only the inspiral phase of binary evolution is considered. 
To include the effect of eccentric orbits, the energy spectrum at the 
$n$ harmonic is expressed as
\be
\label{eq:harmonic}
\frac{{\rm d}E_{{\rm gw},n}}{{\rm d}(nf_{\rm s})}=
\frac{(G\pi)^{\frac{2}{3}}M_{\rm c}^{\frac{5}{3}}}{3f_{\rm s}^{\frac{1}{3}}}
\left(1+\sum_{i=2}^{3} \mathcal{A}_{i} v_{\mathrm{M}}^{i}\right)^{2}
\frac{2}{n}^{\frac{2}{3}}\frac{g(n,e)}{F(e)},
\ee
where the component masses are constrained to $2.5\,M_{\odot} \le m_{2} \le m_{1} \le 100\, M_{\odot}$, with total mass $M_{\rm t} = m_1 + m_2$ and chirp mass $M_{\rm c} = (m_1 m_2)^{3/5} / (m_1 + m_2)^{1/5}$. 
The coefficients $\mathcal{A}_{i}$, where $v_{\mathrm{M}} \equiv (\pi M f_{\mathrm{s}})^{1/3}$, account for post-Newtonian waveform corrections. 
The mass distribution follows the ``FullPop-4.0" model from Ref.~\cite{LIGOScientific:2025pvj}, which is modified from the ``power-law + Dip + Break" model presented in GWTC-3.0~\cite{KAGRA:2021duu}. 
The functions $g(n,e)$ and $F(e)$ are defined in Ref.~\cite{Peters:1963ux}. 
In terms of~\eq{eq:harmonic}, the total energy spectrum at a fixed GW frequency is
\be
\frac{{\rm d}E_{\rm gw}}{{\rm d}f_{\rm s}}
=\sum_{n=1}^{n_{\rm max}}
\frac{{\rm d}E_{{\rm gw},n}}{{\rm d}(nf_{\rm s})},
\ee
where we adopt $n_{\rm max}=1000$ in this work. 

To compute the energy spectrum, the orbital eccentricity evolution must be tracked over the inspiral of binaries. 
For an initial orbital frequency $f_0$\footnote{The orbital frequency is half of the GW frequency.} and eccentricity $e_0$, the frequency-eccentricity relation is given by~\cite{Enoki:2006kj,Huerta:2015pva}:
\be
\frac{f}{f_{0}}=\frac{\mathcal{E}(e)}{\mathcal{E}(e_{0})},
\ee 
with
\be
\mathcal{E}(e)=\frac{\left(1-e^{2}\right)^{3/2}}{e^{18/19}}
\left(1+\frac{121}{304}e^{2}\right)^{-1305/2299}.
\ee

\begin{figure}[h]
	\centering
	\includegraphics[height=6.2cm]{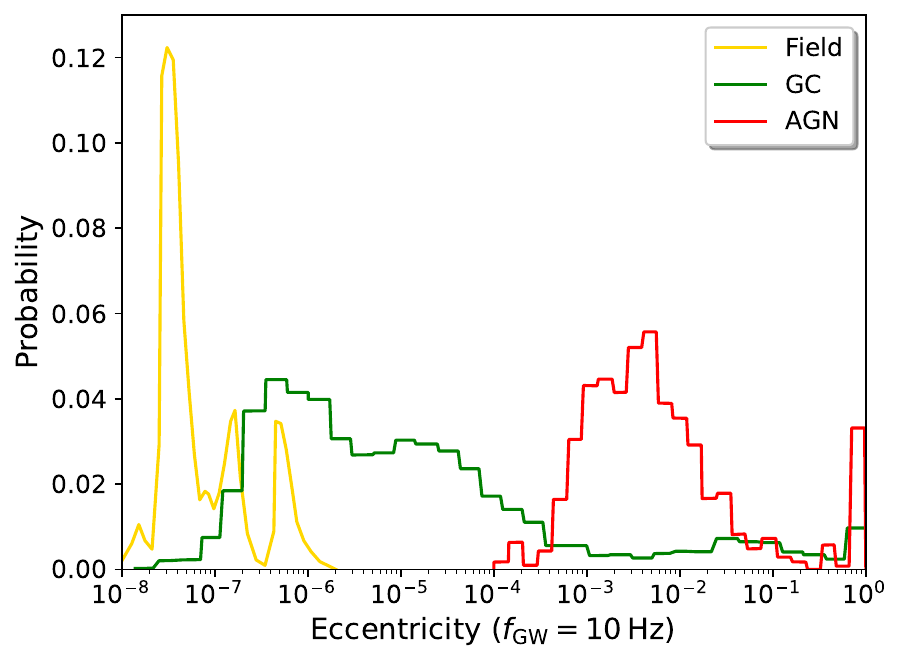}
	\caption{Probability distribution of orbital eccentricity for three different formation channels at an initial GW frequency $f_{\rm GW}$ of $10\,\rm Hz$. The gold, green, and red lines correspond to the field, GC, and AGN cases, respectively.}
	\label{fig:p_e}
\end{figure}

\begin{figure}[h]
	\centering
	\includegraphics[height=6.2cm]{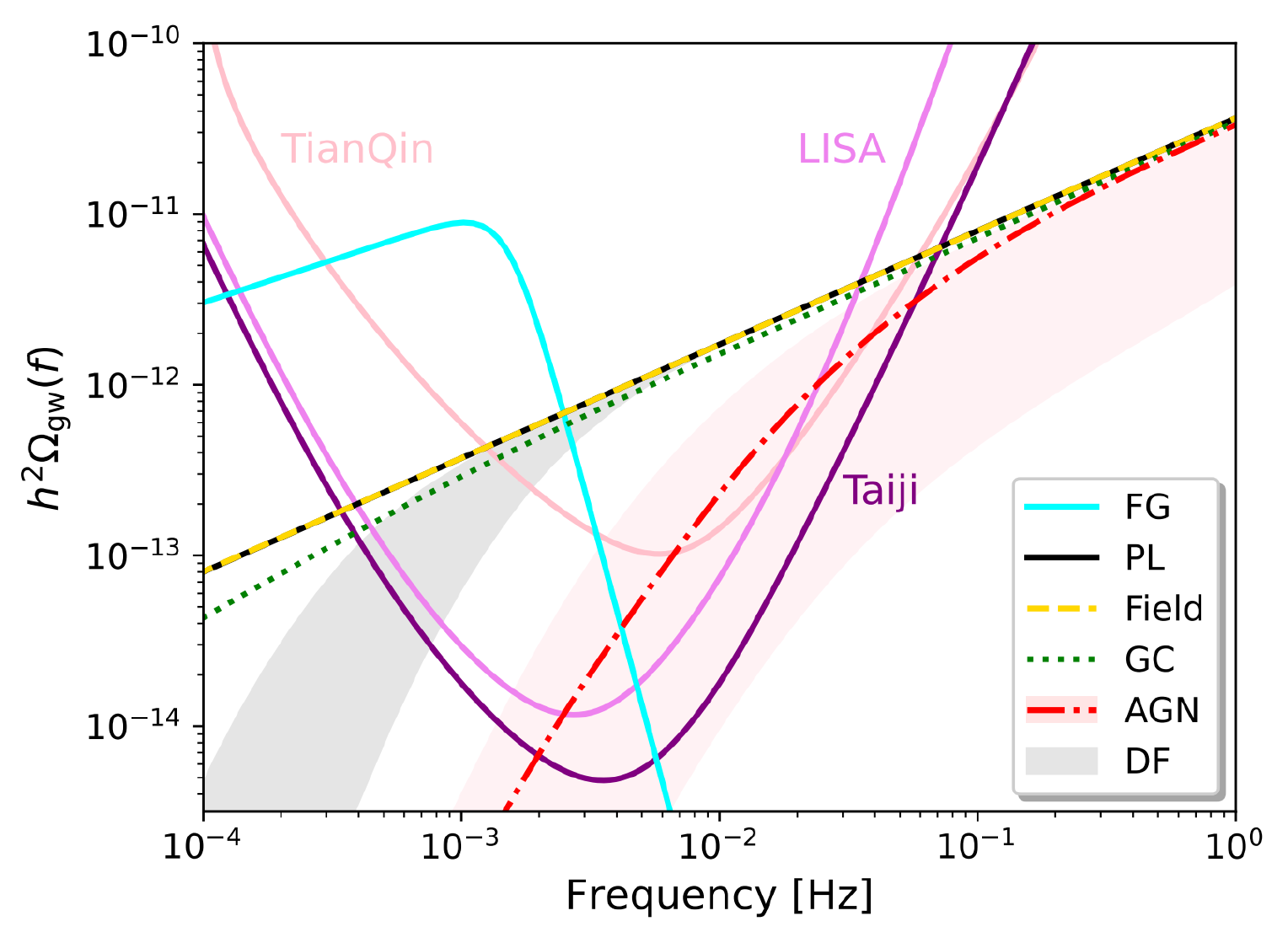}
	\caption{Energy spectral density $\Omega_{\rm gw}$ of the Galactic foreground and the SGWB from different SBBH populations, compared with the PLIS curves for TianQin, LISA, and Taiji for a four-year operation period at $\rho_{\rm th}=1$. The cyan solid line represents the Galactic foreground (FG). The black solid, gold dashed, green dotted, and red dash-dotted lines correspond to the strictly power-law background and the eccentric backgrounds from the field, GC, and AGN cases, respectively. For reference, the pink-shaded region indicates the uncertainty range in $\Omega_{\rm gw}$ due to variations in the eccentricity distribution for the AGN case, while the gray-shaded region represents the range of $\Omega_{\rm gw}$ associated with dynamical friction (DF) under different disk densities.}
	\label{fig:BBH_ecc_model}
\end{figure}

In this paper, we examine three eccentricity models corresponding to different formation channels:
(i) field case: \acp{SBBH} originating from isolated binary evolution in the Galactic field\footnote{For simplicity, SBBH mergers from isolated triples are not considered~\cite{Perets:2012yx,Silsbee:2016djf,Antonini:2017ash,Dorozsmai:2025jlu}.}~\cite{Kremer:2018cir}; 
(ii) GC case: \acp{SBBH} formed in \acp{GC}, including primordial, ejected, in-cluster, and capture binaries~\cite{Rodriguez:2018pss}; 
(iii) AGN case: \acp{SBBH} residing in \acp{AGN}, treated as the fiducial model in~Ref.~\cite{Tagawa:2020jnc}. 
The eccentricity distributions for these models at an initial GW frequency $f_{\rm GW}=10\,\rm Hz$ are shown in Fig.~\ref{fig:p_e}. 
For the field and GC cases, eccentricities are generally below $10^{-3}$. 
In contrast, the AGN case exhibits eccentricities predominantly between $10^{-3}$ and $10^{-2}$, with a non-negligible fraction exceeding. 
This eccentricity distribution suggests that the resulting background could deviate noticeably from a strictly power-law form. 
The energy spectral density also depends on the local merger rate $R_{0}$, which is quoted here with its 90\% central credible interval. 
For clarity, \fig{fig:BBH_ecc_model} illustrates the energy spectral densities computed using the median merger rate of $19\,{\rm{Gpc^{-3}\,yr^{-1}}}$, under the assumption that all SBBHs form through a single case. 
The field case yields an \ac{SGWB} that closely follows a strictly power-law form, with an energy spectral density of $\Omega_{\mathrm{gw}} \approx 8.1^{+3.0}_{-2.1} \times 10^{-13}$ at a reference frequency of $f_{\mathrm{ref}} = 1\,\rm{mHz}$. 
The GC case shows a mild deviation from this trend, while the AGN case, which is characterized by systematically higher orbital eccentricities, results in a strongly suppressed SGWB below $0.1\,\rm Hz$, with amplitude differences reaching several orders of magnitude near $1\,\rm mHz$. 
In addition to the fiducial eccentricity distribution from Ref.~\cite{Tagawa:2020jnc}, we also calculate the energy spectral density under other eccentricity distributions and display their intensity ranges in the pink-shaded region of \fig{fig:BBH_ecc_model}. For certain eccentricity distributions where the eccentricities predominantly exceed 0.1, the SGWB can be strongly suppressed even at frequencies well above $1\,\rm Hz$.

To quantitatively capture deviations from the canonical $f^{2/3}$ scaling, we model the energy spectral density using a modified power-law form,
\be
\label{eq:Omega_ast}
\Omega_{\rm gw}(f)=\Omega_{\rm ast}\left(\frac{f}{f_{\rm ref}}\right)^{2/3}\frac{1}{(f/f_{\rm d})^{\beta}+1},
\ee
where $f_{\rm d}$ is a characteristic drop-off frequency and $\beta < 0$ is a spectral tilt parameter. 
The correction term tends toward unity when $f_{\rm d}$ is sufficiently low or $\beta \to 0$, thereby recovering the strictly power-law behavior.

\begin{table}[h]
	\begin{center}
		\caption{SNRs for the eccentric SBBH-generated SGWB, evaluated for TianQin, LISA, and Taiji, assuming a four-year operation time.}
		\label{tab:snr}
		\setlength{\tabcolsep}{3mm}
		\renewcommand\arraystretch{1.5}
		\begin{tabular}{*{4}{c}}
			\hline
			\hline
			& Field  & GC    & AGN  \\
			\hline 
            TianQin   & 12.9   & 11.2  & 1.6   \\
			LISA      & 64.3   & 55.2  & 3.2   \\
			Taiji     & 171.2  & 149.9 & 12.7   \\
			\hline
			\hline
		\end{tabular}
	\end{center}
\end{table}

Assuming an operation time of four years, we evaluate the expected SNR for the SGWB using the TianQin, LISA, and Taiji detectors. 
The results, summarized in~\tab{tab:snr}, indicate that TianQin yields a significantly lower SNR compared to both LISA and Taiji. 
It is further illustrated by the \ac{PLIS} curves for the three detectors\footnote{It should be noted that TianQin is planned to operate with a half-time observational duty cycle, effectively reducing its total data collection period to half the nominal mission duration.}. 
In particular, TianQin exhibits comparatively weaker sensitivity to the spectral features of SGWB below $0.01\,\rm Hz$, a frequency band that is crucial for distinguishing between SGWBs produced by eccentric SBBHs and the strictly power law background.

It is worth noting that although the observed Galactic foreground may vary across different spaceborne GW detectors due to differences in instrumental noise, LISA and Taiji are expected to observe a nearly identical Galactic foreground~\cite{Liu:2023qap}. 
For simplicity and consistency in our comparative analysis, we therefore adopt a common Galactic foreground model in all subsequent studies and do not explicitly distinguish between the foreground realizations of the detectors. 
\fig{fig:BBH_ecc_model} includes this Galactic foreground corresponding to a four-year observation period, modeled using a broken power law~\cite{Chen:2023zkb,Chen:2024jca},
\be
\Omega_{\rm gw}(f)
=\frac{A_{1}(f/f_{\ast})^{\alpha_{1}}}{1+A_{2}(f/f_{\ast})^{\alpha_{2}}},
\ee 
with parameter values $\{A_{1},A_{2},\alpha_{1},\alpha_{2}\}=\{3.98\times10^{-16},4.79\times10^{-7},-5.7,-6.2\}$. 
The foreground is found to dominate the signal below $1\,\rm mHz$ and remains significant up to approximately $3\,\rm mHz$. 

Furthermore, environmental effects such as dynamical friction and gas accretion typically introduce additional energy dissipation. At low frequencies, these effects can dominate over GW emission and thus suppress the SGWB~\cite{Chen:2025qyj}. 
As an illustration, we consider the dynamical friction, which exerts the strongest influence on the SGWB among environmental effects. The corresponding result is shown as the gray-shaded region in~\fig{fig:BBH_ecc_model}, assuming typical disk densities in the range of $\rho\sim 10^{-11}$-$10^{-8}\,\rm g\,cm^{-3}$. Unlike the AGN case (pink-shaded region), the $\Omega_{\rm gw}$ associated with dynamical friction departs from a strictly power-law form at significantly lower frequencies, where the Galactic foreground already masks it. In light of these considerations, we neglect environmental effects in the remainder of our analysis.

\subsection{Model selection}
Within a Bayesian framework, the task of detecting an SGWB can be approached through model selection. 
A well-recognized challenge in this endeavor arises from the bright Galactic foreground, which is projected to dominate over the instrument noise of spaceborne \ac{GW} detectors. 
This foreground must be carefully accounted for, as it has the potential to significantly bias the inference of other SGWB components~\cite{Bender:1997hs,Nelemans:2001hp,Barack:2004wc,Edlund:2005ye,Ruiter:2007xx,Nelemans:2009hy,Cornish:2017vip,Huang:2020rjf,Liang:2021bde,Staelens:2023xjn,Hofman:2024xar,Liang:2024tgn,Pozzoli:2024wfe}. 
Consequently, this work extends beyond merely confirming the presence of an SGWB above detector noise. 
Instead, we aim to assess the capability of the spaceborne detector to confidently differentiate an SGWB originating from both circular and eccentric SBBH populations against the Galactic foreground. 
To this end, we consider the following three competing hypotheses: 
\begin{itemize}
	\item $\mathcal{H}_{0}$: the signal is due solely to the Galactic foreground;
	\item $\mathcal{H}_{1}$: the signal consists of the Galactic foreground plus a standard power-law SGWB; 
	\item $\mathcal{H}_{2}$: the signal comprises the Galactic foreground and a generalized SGWB described by a corrected power-law model. 
\end{itemize}
The logarithmic Bayes factor $\ln (\mathcal{B}_{10})$, comparing $\mathcal{H}_{1}$ to $\mathcal{H}_{0}$, quantifies the evidence for the presence of any SGWB component beyond the Galactic foreground. 
Additionally, the logarithmic Bayes factor $\ln \mathcal{B}_{21}$, comparing $\mathcal{H}_{2}$ to $\mathcal{H}_{1}$, allows us to assess whether the detected SGWB exhibits spectral features that deviate significantly from a standard power law. 
We now specify the parametrizations adopted for each model. 
The Galactic foreground is modeled using four parameters: two amplitudes $A_{1}$ and $A_{2}$, and two spectral indices $\alpha_{1}$ and $\alpha_{2}$. 
For the SGWB component, since our focus lies in characterizing deviations from the standard power-law form, we fix the spectral index to the theoretically expected value of $2/3$. 
The remaining free parameters for the SGWB are therefore the amplitude $\Omega_{\rm ast}$, the dropoff frequency $f_{\rm d}$, and the deviation index $\beta$. 
In choosing priors for the parameters, we aim to balance weakly constrained ranges with realistic observational bounds to ensure a robust analysis. 
For the parameters of the detector noise and Galactic foreground, we adopt uniform priors on $\log(S_{\rm p,a})$, $\log(A_{1,2})$, and $\alpha_{1,2}$, spanning a range of two units above and below their true values. 
For the parameters of \ac{SGWB}, a Gaussian prior is placed on $\log(\Omega_{\rm ast})$, corresponding to a 90\% credible interval of $[-12.2,-12.0]$\footnote{It corresponds to $\Omega_{\rm ast}=8.1^{+3.0}_{-2.1} \times 10^{-13}$.} at a reference frequency of $1\,\rm mHz$. 
Uniform priors over the range $[-3,0]$ are adopted for both $f_{\rm d}$ and $\beta$.

\begin{figure*}[ht]
	\centering
	\includegraphics[height=8cm]{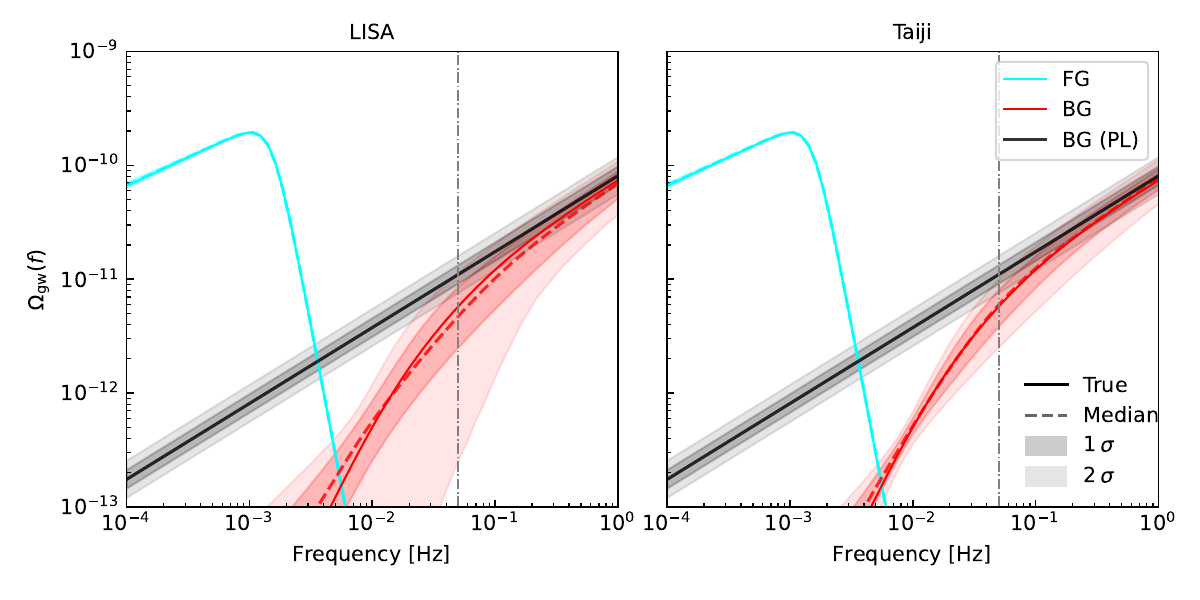}
	\caption{Reconstructed energy spectral density $\Omega_{\rm gw}$ from four-year simulated datasets for LISA (left panel) and Taiji (right panel). The cyan and red lines correspond to the four-year Galactic foreground and the SGWB from SBBHs in AGN, respectively. For comparison, a strictly power-law background is also shown as a black line. The posterior distributions (shaded 1-$\sigma$ and 2-$\sigma$ regions) and their medians (dashed lines) are compared against the true values (solid lines), illustrating the frequency-dependent reconstruction accuracy of each detector. A cutoff frequency of $0.05\,\rm Hz$ is marked by the gray dash-dotted line for reference.}
	\label{fig:PE_tot}
\end{figure*}

\begin{table}
	\begin{center}
		\caption{Logarithm of the Bayes factor, $\ln (\mathcal{B}_{10})$, comparing hypothesis $\mathcal{H}_{0}$ (presence of SGWB) against $\mathcal{H}_{1}$ (absence of SGWB), with 1-$\sigma$ credible intervals.}
		\label{tab:lnB10}
		\setlength{\tabcolsep}{2.5mm}
		\renewcommand\arraystretch{1.5}
		\begin{tabular}{*{4}{c}}
			\hline
			\hline
			& Field   & GC      & AGN  \\
			\hline
            TianQin    & $15.6^{+0.2}_{-0.3}$ & $12.9^{+0.2}_{-0.4}$ & $-1.6^{+0.3}_{-0.5}$\\
			LISA       & $298.2^{+0.5}_{-0.2}$ & $228.9^{+0.3}_{-0.5}$ & $-28.8^{+0.5}_{-0.3}$  \\
			Taiji      & $1903.4^{+0.3}_{-0.4}$ & $1478.0^{+0.4}_{-0.3}$ & $-62.7^{+5.6}_{-4.2}$  \\
			\hline
			\hline
		\end{tabular}
	\end{center}
\end{table}

\begin{table}
	\begin{center}
		\caption{Logarithm of the Bayes factor, $\ln (\mathcal{B}_{21})$, evaluating the relative evidence for an eccentric SBBH-generated SGWB $\mathcal{H}_{2}$ versus a strictly power-law SGWB $\mathcal{H}_{1}$. The 1-$\sigma$ credible intervals are indicated.}
		\label{tab:lnB21}
		\setlength{\tabcolsep}{2.5mm}
		\renewcommand\arraystretch{1.5}
		\begin{tabular}{*{4}{c}}
			\hline
			\hline
			           & Field   & GC      & AGN  \\
			\hline
            TianQin    & $-1.9^{+0.3}_{-0.2}$ & $-1.8^{+0.4}_{-0.2}$ & $2.0^{+0.3}_{-0.3}$\\
			LISA       & $-3.9^{+0.3}_{-0.4}$ & $-3.1^{+0.4}_{-0.2}$ & $30.1^{+0.2}_{-0.6}$  \\
			Taiji      & $-4.8^{+0.4}_{-0.5}$ & $-4.4^{+0.3}_{-0.3}$ & $82.2^{+4.4}_{-5.5}$  \\
			\hline
			\hline
		\end{tabular}
	\end{center}
\end{table}

Following the methodology outlined in Sec.~\ref{sec:method}, we produce four-year simulated data realizations based on autocorrelations of the $A$, $E$, and $T$ channels for TianQin, LISA, and Taiji. 
The dataset is segmented into 11680 intervals, each spanning three hours. 
This segmentation strategy can reduce the number of frequency points by approximately 4 orders of magnitude, while ensuring that the minimum frequency in the Fourier domain remains below $0.1\,\rm mHz$, thereby encompassing the most sensitive frequency band of spaceborne GW detectors. 
In addition, since detector sensitivity degrades significantly at higher frequencies, we apply a cutoff at $0.05\,\rm Hz$ to reduce data volume and enhance computational efficiency.

The resulting logarithmic Bayes factors, $\ln(\mathcal{B}_{10})$ and $\ln(\mathcal{B}_{21})$, are summarized in Tables~\ref{tab:lnB10} and~\ref{tab:lnB21}, respectively. 
In the field and GC cases, TianQin, LISA, and Taiji provide strong evidence for the detection of an SGWB, with values of $\ln(\mathcal{B}_{10})$ significantly exceeding 5, even after accounting for the Galactic foreground. 
Nevertheless, the resulting SGWBs are indistinguishable from a strictly power-law background as indicated by negative values of $\ln(\mathcal{B}_{21})$, and likewise from each other. 
In the AGN case, by contrast, the corresponding SGWB is poorly described by a strictly power-law model, as reflected by the negative values of $\ln(\mathcal{B}{10})$. 
Moreover, $\ln(\mathcal{B}{10})$ exhibits a pronounced downward trend with increasing detection SNR. 
The modified power-law model can offer a better fit in this case, with $\ln(\mathcal{B}_{21})$ values for LISA and Taiji well above 5. 
We now proceed to parameter estimation for this SGWB using LISA and Taiji.

\subsection{Parameter estimation}
Based on the simulated dataset used for model selection, we marginalize over the parameters characterizing both the Galactic foreground and the SGWB component to reconstruct the energy spectral density $\Omega_{\rm gw}$ in a physically meaningful way.

The reconstructed $\Omega_{\rm gw}$ with credible intervals is presented in \fig{fig:PE_tot}. 
At frequencies near $1\,\rm mHz$, where the Galactic foreground reaches its maximum intensity, LISA and Taiji demonstrate comparable sensitivity.
As a result, both detectors recover the foreground component with similar precision, producing posterior estimates that align closely with the injected values and exhibit narrow uncertainties. 
At higher frequencies, where the foreground diminishes rapidly and the SGWB becomes dominant, Taiji's improved sensitivity yields more accurate estimates of the SGWB compared to LISA. 
Furthermore, the reconstruction of the spectrum exhibits significant frequency-dependent variation. 
Below $0.1\,\rm Hz$, the reconstructed intensity deviates more noticeably from the true values. 
This discrepancy can be attributed to two main factors: contamination from the bright Galactic foreground and detector noise at low frequencies, combined with limitations of the simplified spectral model adopted in our analysis. 
Implementing a more flexible parametrization could potentially improve reconstruction accuracy in this band. 
For example, one can incorporate additional spectral shape parameters beyond $f_{\rm d}$ and $\beta$, or adopt a polynomial expansion similar to those used in Galactic foreground modeling in Refs.~\cite{Huang:2020rjf,Liang:2021bde}. 
As the frequency increases, the Galactic foreground drops below the SGWB while detection sensitivity remains high, enabling a more faithful recovery of the SGWB. 
Given that the cutoff frequency is set at $0.05\,\rm Hz$, the reconstruction for $\Omega_{\rm gw}$ is confined to assessing the spectral shape parameters $f_{\rm d}$ and $\beta$. 
To further evaluate the amplitude parameter $\Omega_{\rm ast}$, we extend the reconstruction beyond $0.05\,\rm Hz$. 
The posterior medians are found to remain unbiased and closely follow the injected values, confirming the robustness of the $\Omega_{\mathrm{ast}}$ estimation. 
For comparison, we also show the strictly power-law background. 
Below $0.05\,\rm Hz$, the median reconstructed $\Omega_{\rm{gw}}$ lies entirely outside the 2-$\sigma$ region of the strictly power-law background. 
As the frequency increases, the 1-$\sigma$ and 2-$\sigma$  uncertainty regions of the reconstructed $\Omega_{\rm{gw}}$ progressively narrow and become consistent with those of the strictly power-law background (though the corresponding frequency band above $1\,\rm Hz$ is not shown in Fig.~\ref{fig:PE_tot}).

\section{Summary}\label{sec:Summary}
In this paper, we have investigated the detectability of the SGWB generated by eccentric SBBHs using TianQin, LISA, and Taiji. 
We began by computing the energy spectral density for three distinct SBBH populations: isolated evolved binaries, those in GCs, and those in AGNs. 
Subsequent evaluations of detection SNR and sensitivity revealed that TianQin achieves significantly lower SNR and exhibits poorer sensitivity to the spectral feature of SGWB compared to the other two detectors.

Furthermore, within a Bayesian framework, we first derived a simplified likelihood function that can reduce the number of frequency points requiring processing by 4 orders of magnitude, substantially accelerating data analysis. 
Using this likelihood function, we performed model selection and parameter estimation to assess the capability of detectors to identify an eccentric-SBBH-induced SGWB in the presence of a Galactic foreground. 
Our results indicated that the SGWB from isolated SBBHs and those in GCs are indistinguishable from the strictly power-law background. 
In contrast, the SGWB from binaries in a AGN exhibits marked spectral deviations due to high residual orbital eccentricity, allowing for a clear distinction from the strictly power-law background. 
For the AGN case, we further quantified the precision with which parameters can be constrained using LISA and Taiji. 
The reconstructed SGWB spectrum shows strong frequency-dependent behavior: larger deviations occur below $0.01\,\rm Hz$, reliable extraction is achieved between $0.01$ and $0.05\,\rm Hz$, and the amplitude parameter $\Omega_{\rm ast}$ remains consistently well constrained at higher frequencies.

Two important limitations of our current work should be acknowledged. 
First, our analysis has considered the contributions of an SGWB from individual SBBH populations separately, whereas the observed background will likely comprise a superposition of multiple populations. 
This integration would alter the effective eccentricity distribution of the binaries and consequently modify the spectral transition features of the composite SGWB. 
For unambiguous detection of eccentricity signatures, these spectral transitions must occur within the sensitive frequency bands of the detectors.

Second, our implementation of the null-channel method has assumed perfect knowledge of detector noise properties. 
However, as demonstrated in~\citet{Muratore:2023gxh}, accounting for noise uncertainties can substantially degrade measurement precision, potentially requiring the background amplitude to be around 50 times higher to achieve comparable detection confidence. 
This inherent limitation naturally directs attention toward an alternative method. 
Adopting the cross-correlation method with detector networks emerges as particularly advantageous in this context, as it inherently circumvents the need for precise noise property characterization~\cite{Orlando:2020oko,Seto:2020mfd,Wang:2021njt,Wang:2022pav,Liang:2023fdf,Wang:2023ltz,Zhao:2024yau,Liang:2024tgn}. 
Given the anticipated operational overlap between the LISA and Taiji missions, the LISA + Taiji network can be employed to detect the SGWB from eccentric SBBHs.

\begin{acknowledgments}
This work has been supported by the National Key Research and Development Program of China (No. 2023YFC2206704 and No. 2020YFC2201401), and the Natural Science Foundation of China (Grant No. 12173104). Z.-C.  L. is supported by the Guangdong Basic and Applied Basic Research Foundation (Grant No. 2023A1515111184).
\end{acknowledgments}





\normalem
\bibliographystyle{apsrev4-1}
\bibliography{ref}

\end{document}